\documentclass[12pt]{iopart}
\usepackage{graphicx}
\begin{document}

\title[
{\footnotesize Plasma Production by Helicon Waves with Single Mode Number in Low Magnetic Fields}
]{
Plasma Production by Helicon Waves with Single Mode Number in Low Magnetic Fields
}

\author{G. Sato
\footnote[1]{genta@ecei.tohoku.ac.jp}
, W. Oohara, and R. Hatakeyama }

\address{Department of Electronic Engineering, Tohoku University, Sendai 980-8579, JAPAN}

\begin{abstract}
Radio-frequency discharges are performed in low magnetic fields (0-10 mT) 
using three types of helicon-wave exciting antennas with the azimuthal mode number of $|m|$ = 1.
The most pronounced peak of plasma density is generated in the case of phased helical antenna
at only a few mT,
where the helicon wave with $|m| = 1$ is  purely excited and propagates. 
An analysis based on the dispersion relation well explains the density-peak phenomenon 
in terms of the correspondence between the antenna one-wavelength and the helicon wavelength.
The $m=+1$ helicon wave propagates even in high magnetic fields where the density peaks are not observed, 
but the $m=-1$ helicon wave disappers. 
It is expected theoretically that the $m=-1$ helicon wave shows cutoff behavior in a low density region,
and the cutoff of $m=-1$ helicon wave experimentally observed coincides with the calculated one.
\end{abstract}




\section{Introduction}

The helicon-wave discharge has been extensively investigated 
because a high electron density ($10^{12} - 10^{13}$ cm$^{-3}$) is obtained with comparatively ease 
under a low gas pressure of a few hundreds mPa [1,2], 
being used as plasma source for applications to material processings [3,4], 
such as etching, plasma chemical vapor deposition, etc. 
Conventional helicon-wave discharges need strong magnetic fields ($B_0 \gg 10$ mT) 
generated by electromagnets with a measurable amount of power, 
which has commercially a disadvantage for operating helicon-wave plasma sources with the lower running cost. 
For the purpose of expanding application fields of the helicon-wave source, 
it is significant to  make the best of an efficient plasma production 
in a very low magnetic field.
In several previous works concerning the mechanism of the plasma production 
with anomalously high density [5-7], on the other hand, 
a density increment enhanced by the helicon waves is observed
under a weak magnetic field much smaller than the magnetic fields 
that a mode transition from an inductive-coupled discharge to the helicon-wave discharge appears. 
However, the reason why the helicon waves can be excited within narrow range of low magnetic fields 
has not been exhibited at all. 
Chen [8] has recently suggested using a numerical code that the density peak observed in low magnetic fields is 
caused by an increase in the plasma loading, 
which is due to the reflection of electromagnetic waves at an endplate near the antenna.
This effect is noticeable only in the case of a loop antenna with the azimuthal mode number $m = 0$,  
while the density peak is also observed in other experiments using the $|m|=1$ antennas. 
Therefore it is required to systematically investigate the density peak mechanism, 
which leads to the development of the efficient plasma source in low magnetic fields.\\
\indent
On the basis of the background mentioned above, 
we investigate the $m=+1$ and $m=-1$ helicon wave discharges with low density and low magnetic fields 
by using three types of $|m|=1$ rf antennas, 
a phased multiple helical-antenna (PMH), a half-wavelength helical antenna (HH) [9], 
and a double half-turn antenna (DHT) [5]. 
Since PMH is possible to excite spatiotemporally-rotating electromagnetic fields 
with fixing the wave numbers parallel and perpendicular to magnetic field lines in a plasma, 
the helicon-wave behavior can be clarified even 
in the magnetic field lower than the threshold field, 
above which the conventional helicon wave is excited. 
\\
\section{Experimental Setup}
\indent
The experiment apparatus is shown in Fig. 1, 
which consists of an antenna section of a Pyrex tube (10 cm in diameter and 40 cm in length) 
and a measurement section of a stainless chamber (26 cm in diameter and 89 cm in length). 
An endplate is set at $z=59$ cm from the entrance of the measurement section. 
In order to excite selectively $m=+1$ and $m=-1$ helicon waves,
the phased multiple helical antenna (PMH) is used in this experiment. 
PMH consists of four elements, 
each of which is twisted into helix in the left-hand direction along the $+z$ axis
with the antenna length of $l_z = 20$ cm. 
An rf signal from a signal generator is divided into two, 
the phase difference of them is fixed at 90$^{\circ}$ by a phase shifter.
The rf signals driven with 90$^{\circ}$ phase difference are amplified 
and supplied to two pairs of the antenna elements through the impedance matching circuits, 
each of which is divided into two branches and transmitted to the opposite terminals 
of the two elements spaced azimuthally 180$^{\circ}$. 
Accordingly, the spatiotemporally-rotating (circular-polarized) rf fields with $|m| = 1$ 
are generated by PMH. 
The temporal rotation is fixed in the right-hand direction with respect to the $+z$ direction. 
The uniform magnetic field less than 11 mT is produced along the $z$ axis by solenoidal coils. 
The temporal rotation is fixed in the right-hand direction with respect to the $+z$ direction, 
and the change from $m = +1$ to $m = -1$ is attained 
by reversing the magnetic field direction from $B_0 >$ 0 to $B_0 <$ 0. 
In addition to PMH, HH ($l_z = 20$ cm) and DHT are also used 
as the conventional helicon-wave antennas with $|m|=1$, 
each of which is supplied with the rf power using only one transmission line. 
Typical operating parameters include the rf frequency of $\omega /2\pi = 13.56$ MHz, 
the rf power of $P_{rf} = 1000$ W,
and filling argon pressure of $5 \times 10^{-2}$ Pa. 
Plasma parameters and magnetic fluctuations are measured 
by an axially movable ($z$-axis) Langmuir-probe and a magnetic probe, respectively. 

\section{Results}
\indent
Figures 2(a), (b), and (c) give the electron densities depending on the magnetic field for the several rf powers 
using PMH, HH, and DHT, respectively.
In the case of PMH, 
the electron density increases with increasing in $B_0$ for $m=+1$, 
and the density has a peak at $B_0 \sim +5$ mT for 2000 W,
where the density attains up to one order of magnitude larger than that at $B_0 = 0$ mT, i.e., ICP mode. 
The peak density becomes higher with increasing in $P_{rf}$, 
and $B_0$ yielding the density peak is different between each rf power. 
On the other hand, the density scarcely increases for $m=-1$, 
and the peak density is almost one-fifth the magnitude of that for $m = +1$. 
The density difference for between $m=+1$ and $m=-1$ becomes small in the case of HH (Fig. 2(b)),
and is not observed in the case of DHT as shown in Fig. 2(c).
The reason for the density increment for $B_0 < 0$ is that DHT concurrently excites $m= \pm 1$ fields, 
and $m=+1$ field excited regardless of magnetic-field direction gives rise to the density increment.
Since HH includes double half-turn parts at the both ends and excites $m= \pm 1$ fields,
the density increment is caused in $B_0 < 0$ (Fig. 2(b)), which is larger than that in the case of PMH.
\\
\indent
When the $z$ component of the magnetic fluctuation $B_z$ is measured in $z>0$ cm, 
waves with large amplitude are observed to propagate toward the downstream.
The radial amplitude profiles of them for $B_0 = +3$, $-3$, $+10$, and $-10$ mT 
are plotted in Figs. 3(a), (b), (c), and (d), respectively. 
The solid lines of each figure show the fitted curves expressed by composite function
of the zeroth and first order Bessel functions ($J_0$ and $J_1$).
In the case of $|B_0| = 3$ mT, $B_z$ is formed with $J_1$ approximately for both of $m=+1$ and $m=-1$ modes
and the radial position where the amplitude goes to zero is $r= 5.5$ cm obtained from fitted curves, 
which agrees with the plasma radius determined by the Pyrex tube size. 
The difference in magnitude between the right and left peaks shows that the wave includes $J_0$ component 
about 10 $\%$, 
the boundary of which exists at $r=13$ cm, namely, the stailess chamber wall. 
The generation mechanism of $J_0$ component is not cleared for the moment. 
The amplitude for $B_0 = +10$ mT (Fig. 3(b)) is also fitted to $J_1$ Bessel function approximately. 
The $B_z$ radial profile doesn't depend on $B_0$ for $m=+1$, 
but it is found that the amplitude for $B_0 =-10$ mT is not fitted to $J_1$ but $J_0$ in spite of $m=-1$ excitation,
as shown in Fig. 3(d).
The radial position where the amplitude goes to zero is $r=13$ cm 
as well as $J_0$ components included slightly in $|m|=1$ helicon waves. 
But the amplitude of the $m=0$ wave for $B_0 = -10$ mT is greater 
than that of $J_0$ components for other cases of Figs. 3(a), (b), and (c). 
\\
\indent
The theoretical dispersion relation of the waves is expressed for a cold uniform plasma: 
\begin{eqnarray}
	P n_{z}^4 + (n_{\perp}^2 (S + P) - 2PS)n_{z}^2 
	+ (Sn_{\perp}^2 - RL)(n_{\perp}^2 - P) = 0,
	\label{eq1}
\end{eqnarray}
where $P$, $S$, $R$, and $L$ are followed by Stix notation [10]. 
$n_{z}$ and $n_{\perp }$ are the refraction indexes parallel and perpendicular to $B_0$, respectively.
The perpendicular wave number $k_{\perp}$ is 58 m$^{-1}$ calculated from Figs. 3(a) and 3(b). 
Figure 4 shows the dispersion relations of the $m=+1$ helicon wave 
($k_z$ vs $B_0$), 
where closed circles are measured by a b-dot probe
and solid line is calculated from $n_e$ depending on $B_0$ for PMH at $P_{rf}=1000$ W (Fig.~2(a)).
The measured dispersion relation is consistent with the theoretical curve of the helicon wave
in the range of $B_0 >$ 2 mT, 
while that differs $B_0 < 2$ mT. 
Open circles indicate the maximum amplitude of $B_z$ depending on $B_0$, 
which is extremely small in $B_0 < 2$ mT.
Hence, the discrepancy of the dispersion relation in $B_0 < 2$ mT 
is caused by the influence of antenna near-fields rather than the helicon wave with strong damping. 
\\
\indent
When the electron density and the magnetic field are given, 
the wavelength allowed to exist in the plasma can be calculated from the dispersion relation. 
Thus, the relationships between the electron density $n_e$ and the axial wavelength $\lambda _z$ 
for $P_{RF} = 1000$ W are presented in Fig.~5, 
where $\lambda _z$ is calculated from Fig.~2 and Eq. (1) 
with $\lambda _\perp$ fixed by the radial boundary as mentioned above.
The electron density becomes high at a specific wavelength, i.e.,
$\lambda _z$ = 20 cm (40 cm) for PMH (HH).
Those wavelengths correspond with one axial wavelengths of PMH ($l_z$) and HH ($2 l_z$), respctively. 
Therefore the helicon wave is excited strongly in the plasma 
with the wavelength $\lambda _z$ determined by the electron density and the magnetic field, 
which becomes comparable to the one wavelength of the antenna. 
As shown in Fig.~5, on the other hand, we can recognize no clear-cut relationship between $\lambda _z$ and $n_e$ 
for the DHT case, 
since the electromagnetic field with the fixed wave number of the $z$ component is hardly excited by DHT.
Instead the density is observed to slightly increase around $\lambda _z = 59$ cm 
corresponding with the axial boundary length,
which is considered due to exciting the standing-wave [11].
\\
\indent
In the previous works, 
the characteristics of $|m|=1$ helicon wave in an inhomogeneous plasma have been investigated 
theoretically and experimentally [12,13]. 
Kr\"amer suggested that the $m=-1$ helicon wave in an inhomogeneous plasma is damped more strongly than 
that in a homogeneous plasma, and which has a cutoff in a low density region [13]. 
Considering Gaussian density profiles $n_e = n_0^d \exp(-(r/d)^2)$, 
the cutoff equation is derived
\begin{eqnarray}
	g(r) \approx \alpha^2 - k_z^2 - \frac{2m^2}{d^2} + \frac{2m \alpha}{k_z d^2},
	\label{eq5}
\end{eqnarray}
where $d$ is the profile width yeilding the density of $1/e$ and $k_z$ is the wave number parallel to $B_0$, 
and $\alpha   = {k_0^2}/{k_z} {\omega_{pe}^2}/{\omega \omega_{ce}}$ with $k_0=\omega /c$. 
$g(r) > 0$ is required in the case where the helicon wave exists in an inhomogeneous plasma. 
For the $m=+1$ and $m=0$ modes, $g(r)$ is always positive and the wave can propagate in nonuniform plasmas.
However, the $m=-1$ helicon wave can only propagate in the higher density region above a critical density. 
The critical density curve is theoretically described on the $B_0-n_e$ plane using Eq.(2), as shown in Fig. 6, 
where $d$ is obtained from the radial density profile measured by the Langmuir probe. 
$k_z$ is calculated by using Eq.(\ref{eq1}) 
and the perpendicular wavenumber is $k_{\perp}=$ 69 m$^{-1}$ obtained from Fig. 3(c).
$g(r)$ is constantly positive (negative) in the gray (white) zone, 
which is called propagation (cutoff) region.
Dots in Fig. 6 indicate the electron densities measured, 
which belong to the propagation region only when $1 \le B_0 \le 5$ mT.
It can be said that the $m=0$ helicon wave exists when the background plasma density
gets into the cutoff region of $B_0 >$ 5 mT.
\\
\section{Conclucion}
\indent
In summary, we have clarified a new class of practical aspect of the helicon-wave discharge 
employing various $|m|=1$ antennas
in magnetic fields much lower than the ordinary magnetic field where the density jump occurs.
It is found that the effective excitation of the helicon wave and the efficient plasma production 
causally take place even in the very weak magnetic field 
when the axial wavelength of the phased helical antenna launching spatiotemporally-rotating electromagnetic fields 
is matched to that of the helicon wave determined by the plasma dispersion relation. 
The $m=+1$ helicon wave propagates in the plasma independent of the magnetic field strength. 
On the other hand, 
the $m=-1$ helicon wave propagates only in very low magnetic fields where density peaks are observed. 
The cutoff of $m=-1$ helicon wave is theoretically concluded to exist in the region of $B_0 > 5$ mT,
where the helicon wave propagates with $m=0$.
\\
%

\clearpage
\vspace{20mm}
\begin{figure}[hb]
\begin{center}
\includegraphics[width=100mm,clip]{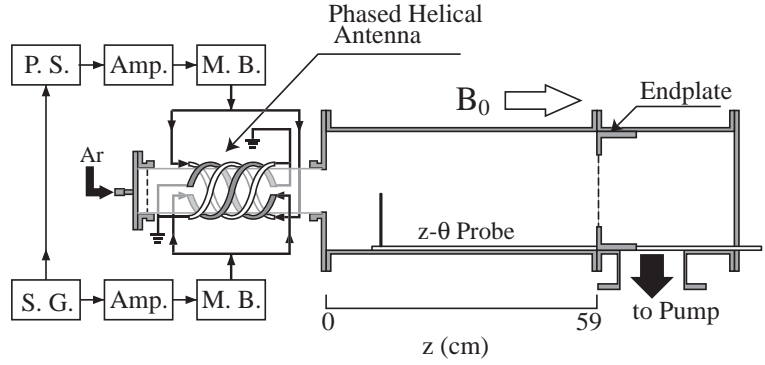}
\caption{\label{fig:epsart} Schematic diagram of experimental setup.}
\end{center}
\end{figure}
\vspace{20mm}
\begin{figure}[ht]
\begin{center}
\includegraphics[width=150mm,clip]{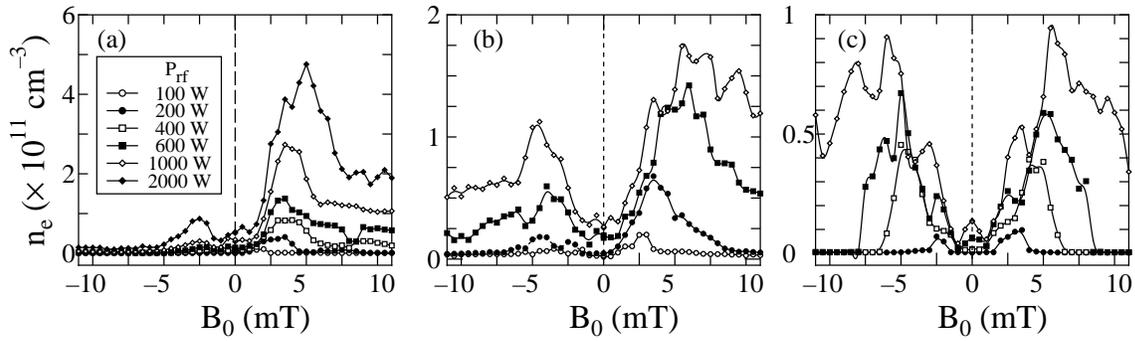}
\end{center}
\caption{\label{fig:2} Electron density depending on magnetic field for various rf powers $P_{rf}$
with (a) phased multiple helical-antenna (PMH), (b) half-turn helical antenna (HH), 
and (c) double half-turn antenna (DHT).
}\end{figure}
\vspace{20mm}
\clearpage
\begin{figure}[hb]
\begin{center}
\includegraphics[width=90mm,clip]{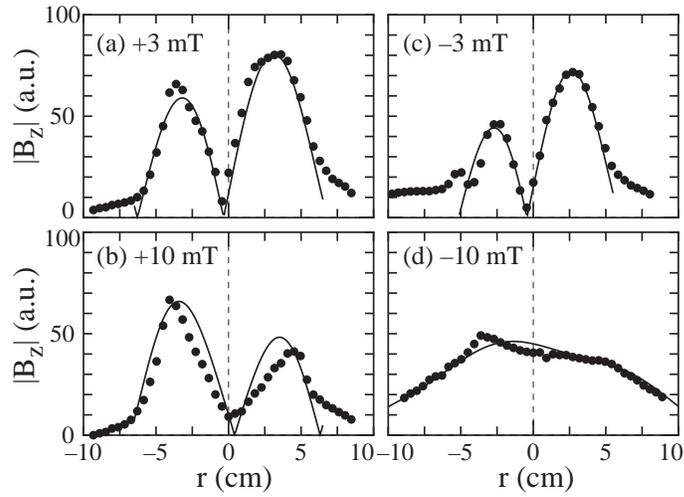}
\caption{\label{fig:epsart}
Radial amplitude profiles of $B_z$ for $B_0 = +3$ mT (a), $+10$ mT (b), $-3$ mT (c), and $-10$ mT (d). 
Dots are measured amplitudes and solid curves are fitted curves based on the Bessel composite functions.
}
\end{center}
\end{figure}
\vspace{20mm}
\begin{figure}[hb]
\begin{center}
\includegraphics[width=70mm,clip]{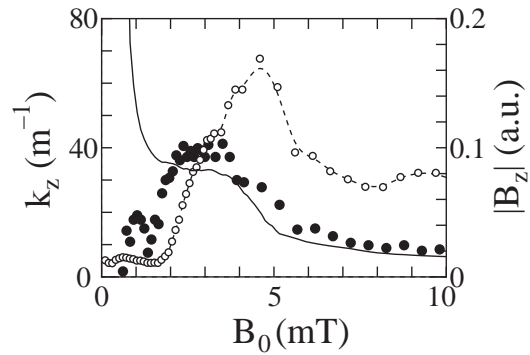}
\end{center}
\caption{\label{fig:epsart}
Dispersion relations of helicon wave measured ($\bullet$) with PMH at $P_{rf}=1000$ W
and calculated (solid line).
The maximal amplitude of axial magnetic fluctuation ($\circ$) has a peak around $B_0 = 5$ mT.
}
\end{figure}
\vspace{20mm}
\clearpage
\begin{figure}[h]
\begin{center}
\includegraphics[width=58mm,clip]{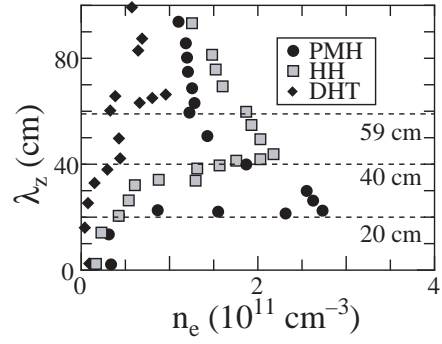}
\end{center}
\caption{\label{fig:4} Relationships between electron density and wavelength 
at $P_{rf} =1000$ W for PMH, HH, and DHT. 
}
\end{figure}
\vspace{20mm}
\begin{figure}[h]
\begin{center}
\includegraphics[width=68mm,clip]{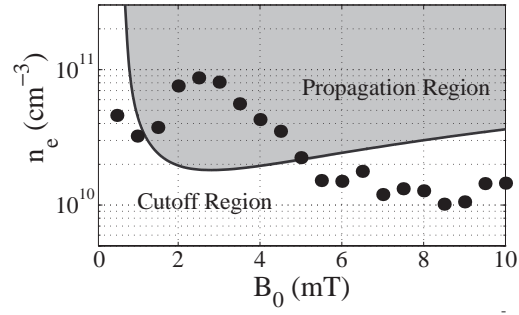}
\caption{\label{fig:6} 
Electron density depending on the magnetic field for $B_0 < 0$ ($\bullet$) 
and theoretical cutoff line of $m=-1$ helicon wave on $B_0 - n_e$ plane  (solid line). 
Upper and lower regions of the solid line indicate the propagation and cutoff regions, respectively.
}
\end{center}
\end{figure}

\end{document}